\documentstyle[twocolumn,twoside,psfig]{article}
\newcommand{\hb}{\\ \hspace*{2ex}}

\begin{document}
\title{IMF AND EVOLUTION OF CLOSE BINARIES AFTER STARFORMATION BURSTS}
\author{S.B.\,Popov$^1$, M.E.\,Prokhorov$^1$, V.M.\, Lipunov$^{1,2}$\\[2mm]
\begin{tabular}{l}
 $^1$ Sternberg Astronomical Insitute\hb
 Universitetskii pr. 13, Moscow 119899 Russia, {\em polar@xray.sai.msu.su}\\
$^2$ Department of physics, Moscow State University, {\em
lipunov@sai.msu.su}\\[2mm]
\end{tabular}
}
\date{}
\maketitle

ABSTRACT.
    This paper is a continuation and development of our previous
articles (Popov et al., 1997, 1998).
We use ``Scenario Machine'' (Lipunov et al., 1996b) -- the population synthesis
simulator (for single binary systems calculations 
the program is available in WWW:
{\bf http://xray.sai.msu.ru/}
(Nazin et al., 1998)) -- 
to calculate evolution of populations of several
types of X-ray sources during the first 20 Myrs after a starformation burst.

We examined the evolution  
of 12 types of X-ray sources in close binary systems (both with neutron
stars and with black holes) for different parameters of the IMF -- slopes: 
$\alpha=1$, $\alpha=1.35$ and $\alpha=2.35$ and
upper mass limits, $M_{up}$: 120 $M_{\odot}$, 60 $M_{\odot}$
and 40 $M_{\odot}$. Results, especially for sources with black holes, are
very sensitive to variations of the IMF, and it should be taken into account
when fitting parameters of starformation bursts.

Results are applied to several regions of recent starformation in
different galaxies: Tol 89, NGC 5253, NGC 3125, He 2-10, NGC 3049. 
Using known ages and total masses of starformation
bursts (Shaerer at al., 1998)
we calculate expected numbers of X-ray sources in close binaries for
different parameters of the IMF. Usually, X-ray transient sources consisting
of a neutron star and a main sequence star are most abundant, but for very
small ages of bursts (less than $\approx 4$ Myrs) 
sources with black holes can become more abundant. 
\\[1mm]
{\bf Key words}: Stars: binary: evolution;\\[2mm]

{\bf 1. Introduction}\\[1mm]

 Theory of stellar evolution and one of the strongest tools 
of that theory -- population synthesis -- 
are now rapidly developing branches of astrophysics.
Very often  only the evolution of single stars is modeled,
but it is well known
that about 50\% of all stars are members of  binary systems,
and a lot of different astrophysical objects are products
of the evolution of binary stars. We argue, that often it is
necessary to take into account the evolution of close binaries
while using the population synthesis in order to avoid serious  errors. 

 Initially this work was 
stimulated by the article Contini et al. (1995),
where the authors suggested an unusual form of the initial mass function
(IMF) for the explanation of the observed properties
of the  galaxy Mrk 712 . They suggested the ``flat'' IMF with the exponent  
$\alpha=1$ instead of the Salpeter's value  $\alpha=2.35$.
Contini et al. (1995) didn't take into account binary systems, so
no words about the influence of such IMF  
on the populations of close binary stars could be said.
Later Shaerer (1996) showed that the observations could be explained
without the IMF with $\alpha=1$. 
Here we try to determine the  influence of the 
variations of the IMF on the evolution of compact binaries
and apply our results to seven regions of starformation (Shaerer et al.,
1998, hereafter SCK98).

Previously (Lipunov et al., 1996a) we used  the
``Scenario Machine'' for  calculations of  populations of
X-- ray sources after a burst of starformation
at the Galactic center. Here, as before in Popov et al.
(1997, 1998), we model a general
situation --- we make calculations for a typical starformation burst.
We show results on 
twelve types of binary sources with significant X-ray luminosity for three
values of the upper mass limit for three values of $\alpha$.\\[2mm]

{\bf 2. Model}\\[1mm]

Monte-Carlo method for statistical simulations of binary evolution
are now widely used in astrophysics:  for analysis of radio pulsar
statistics, for formation  of the galactic   
cataclysmic variables etc.
(see the review in van den Heuvel 1994). 

Monte-Carlo simulations of binary star evolution allows one to
investigate the evolution of a large ensemble of binaries  and to
estimate the number of binaries at different
evolutionary stages. Inevitable simplifications in the
analytical description of the binary evolution that we allow in our
extensive numerical calculations, make those numbers
approximate to a factor of 2-3.  However, the inaccuracy of direct
calculations  giving the numbers of different binary types
in the Galaxy (see e.g. van den Heuvel 1994) 
seems to be comparable to what follows from the
simplifications in the binary evolution treatment.  

 In our analysis of binary evolution, we use the ``Scenario Machine'', a
computer code, that incorporates current scenarios of binary
evolution 
and takes into account the influence of magnetic field of
compact objects on their observational appearance. A detailed description
of the computational techniques and input assumptions is summarized
elsewhere (Lipunov et al. 1996b; see also:
{\bf http://xray.sai.msu.ru/\~ \, mystery/articles/review/}), 
and here we briefly list only principal parameters and initial distributions.

\unitlength=1in
\begin{figure}[t]
\vbox{\psfig{figure=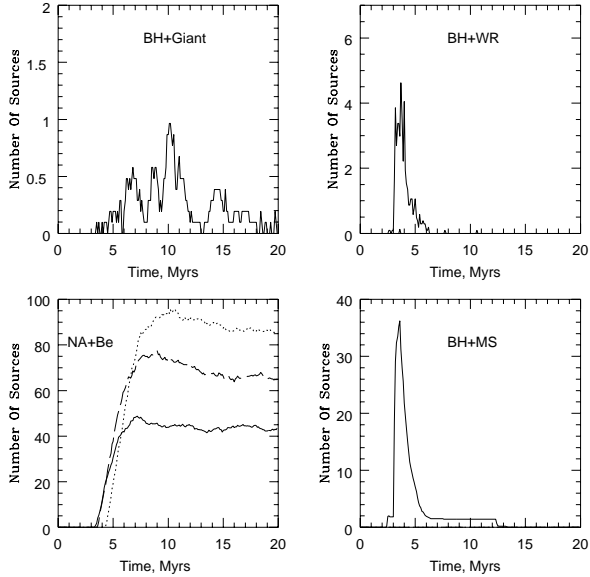,width=3.26in}}
%\begin{picture}(3.26,3.14)
%\vspace{3.14in}
%\put(3.26,3.14){figure1.eps}
%\vbox{\psfig{figure=figure1.eps, width=3.26in}}
%\end{picture}
\caption{Evolution of numbers of binary systems after a burst of   
starformation. $\alpha=1.35$.
BH+Giant -- A BH with a He-core Star (Giant).
BH+WR -- A BH with a Wolf--Rayet Star. 
NA+Be -- An Accreting NS with a Main Sequence Star
                (Be-transient).
BH+MS --  A BH with a Main Sequence Star}
\end{figure}

\unitlength=1in
\begin{figure}[h]
\vbox{\psfig{figure=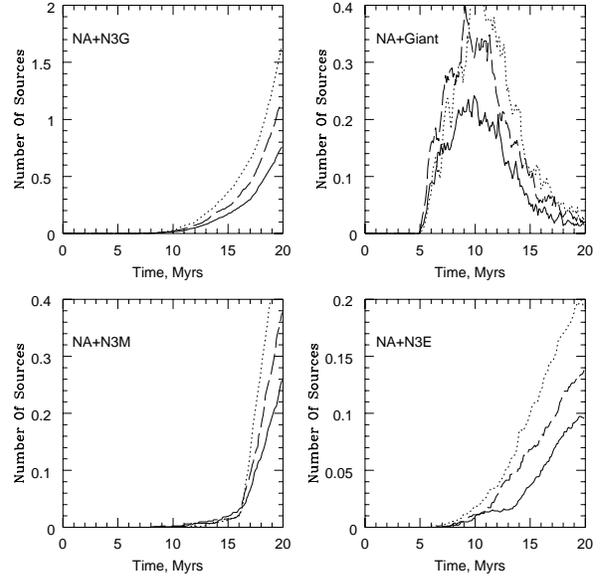,width=3.26in}}
%\begin{picture}(3.26,3.14)
%\vspace{3.14in}
%\put(3.26,3.11){figure3.eps}
%\end{picture}
\caption{Evolution of numbers of binary systems after a burst of
starformation. $\alpha=1.35$.
NA+N3G --  An Accreting NS with a Roche-lobe filling
                  star, when the binary loses angular momentum due to
                   gravitational radiation.
NA+Giant --  An Accreting NS with a He-core Star (Giant).
NA+N3M -- An Accreting NS with a Roche-lobe filling  
                  star, when the binary loses angular
                  momentum due to magnetic wind.
NA+N3E --An Accreting NS with a Roche-lobe filling star
(nuclear evolution time scale).}
\end{figure}

We trace the evolution of binary systems during the first 20 Myrs after
their formation in a starformation burst. Obviously, only stars that are
massive enough (with masses $\ge 8-10~ {\rm M}_\odot$) can evolve off
the main sequence during the time as short as this to yield compact
remnants: neutron stars (NSs) and black holes (BHs).
Therefore we consider only massive binaries, i.e. those having the mass of
the primary (more massive) component in the range of $10 \, {\rm M}_\odot$--
$M_{up}$.

We assume that a NS with a mass of $1.4~ {\rm M}_{\odot}$ is formed
as a result of the collapse of a star, whose core mass prior to collapse was
$M_*\sim (2.5-35)~{\rm M}_{\odot}$. This corresponds to an initial mass 
range $\sim (10 - 60)~{\rm M}_{\odot}$, taking into account that a massive
star can lose more than $\sim (10-20)\%$ of its initial mass during the   
evolution with a strong stellar wind.
The most massive stars are assumed to collapse into a BH once
their mass before the collapse is $M>M_{cr}=35~ {\rm M}_\odot$. 
The BH mass is
calculated as $M_{bh}=k_{bh}M_{cr}$, where the parameter $k_{bh}$ is
taken to be 0.7.

The mass limit for NS (the Oppenheimer-Volkoff limit) is taken to be
$M_{OV}=2.5~ {\rm M}_\odot$, which corresponds to a hard equation of
state of the NS matter.

We made calculations for several values of the coefficient $\alpha$:

$$
   \frac{dN}{dM} \propto M^{-\alpha}
$$

We calculated $10^7$ systems in every run of the program.
Then the results were normalized to the total mass of binary stars 
in the starformation burst.
We also used different values of the upper mass limit, $M_{up}$.\\[2mm]

\unitlength=1in
\begin{figure}[h]
\vbox{\psfig{figure=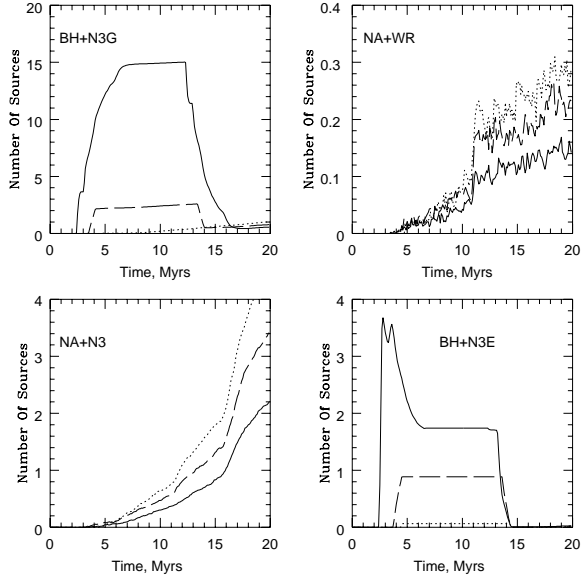,width=3.26in}}
%\begin{picture}(3.26,3.14)
%\vspace{3.14in}
%\put(3.26,3.14){figure2.eps}
%\end{picture}
\caption{Evolution of numbers of binary systems after a burst of
starformation. $\alpha=1.35$.
BH+N3G --  A BH with a Roche-lobe filling star, when the
binary loses angular momentum by gravitational radiation.
NA+WR -- An Accreting NS with a Wolf--Rayet Star. NA+N3 -- An Accreting NS
with a Roche-lobe filling star
               (fast mass transfer from the more massive star).
BH+N3E -- A BH with a Roche-lobe filling star (nuclear
evolution time scale).}
\end{figure}

{\bf 3. Results}\\[1mm]

 On the figures we show some of the results of our calculations
(full results can be found in the electronic preprint (Popov et al. 1999)).
On all graphs on the X- axis we show the time after the 
starformation burst in Myrs, on the Y- axis --- number of 
the  sources of the selected type that exist at the particular moment.

\noindent
On the figures results are shown for three values 
of upper mass limits: $120 M_{\odot}$ -- solid lines,
$60 M_{\odot}$ -- dashed lines, $40 M_{\odot}$ -- dotted lines.

 The calculated numbers were normalized for $10^6 
\, M_{\odot}$ in binary stars. We show on the figures
and in tables
only systems with the luminosity of compact object greater than $10^{33}\,
{\rm erg/s}$.

\begin{table}[h]
\caption[]{He 2-10;  age 5.5 Myrs; total mass $10^{6.8} M_{\odot}$}
\begin{tabular}{|l||c|c|c|c|c|c|}
\hline
 Slope&2.35&2.35&2.35& 1.35&1.35&1.35\\
\hline     

Up.mas.&120&  60&  40&   120&  60&  40  \\

%\hline
\hline
bh+ms & 0 &  0 &  0 &   16 &  0 &  0 \\  

bh+giant & 0 &  0 &  0 &    0 &  0 &  0 \\

bh+n3e& 1 &  0 &  0 &    9 &  4 &  0 \\

bh+n3g& 4 &  1 &  0 &   62 & 10 &  0 \\

bh+wr & 0 &  0 &  0 &    1 &  0 &  0 \\

na+ms &24 & 22 & 15 &  187 &241 &165 \\

na+n3 & 0 &  0 &  0 &    0 &  0 &  0 \\

na+wr & 0 &  0 &  0 &    0 &  0 &  0 \\

na+n3m& 0 &  0 &  0 &    0 &  0 &  0 \\

na+n3e& 0 &  0 &  0 &    0 &  0 &  0 \\

na+n3g& 0 &  0 &  0 &    0 &  0 &  0 \\

na+giant & 0 &  0 &  0 &    0 &  0 &  0 \\

\hline
\end{tabular}
\end{table}

\begin{table}[h]
\caption[]{He 2-10;  age 6.0 Myrs; total mass $10^{6.8} M_{\odot}$}
\begin{tabular}{|l||c|c|c|c|c|c|}
\hline
 Slope&2.35&2.35&2.35& 1.35&1.35&1.35 \\
\hline     

Up.mas.&120&  60&  40&   120&  60&  40  \\

%\hline
\hline
bh+ms &  0&  0 &  0 &   9  & 0  & 0  \\

bh+giant &  0&  0 &  0 &   1  & 0  & 0  \\

bh+n3e&  1&  0 &  0 &   9  & 4  & 0  \\

bh+n3g&  4&  1 &  0 &  65  &11  & 0  \\

bh+wr &  0&  0 &  0 &   0  & 0  & 0  \\

na+ms & 29& 30 & 22 & 198  &283 &233 \\

na+n3 &  0&  0 &  0 &   0  & 1  & 1  \\

na+wr &  0&  0 &  0 &   0  & 0  & 0  \\

na+n3m&  0&  0 &  0 &   0  & 0  & 0  \\

na+n3e&  0&  0 &  0 &   0  & 0  & 0  \\

na+n3g&  0&  0 &  0 &   0  & 0  & 0  \\

na+giant &  0&  0 &  0 &   0  & 1  & 0  \\

\hline
\end{tabular}
\end{table}

\begin{table}[h]
\caption[]{NGC5253A; age 3.0 Myrs; total mass $10^{6.6} M_{\odot}$}
\begin{tabular}{|l||c|c|c|c|c|c|}
\hline
 Slope&2.35&2.35&2.35& 1.35&1.35&1.35\\
\hline

Up.mas.&120&  60&  40&   120&  60&  40\\

\hline     
%\hline
bh+ms & 0 &  0 &  0 &    5 &  0 &  0 \\

bh+giant & 0 &  0 &  0 &    0 &  0 &  0 \\

bh+n3e& 1 &  0 &  0 &   10 &  0 &  0 \\

bh+n3g& 1 &  0 &  0 &   11 &  0 &  0 \\

bh+wr & 0 &  0 &  0 &    0 &  0 &  0 \\

na+ms & 0 &  0 &  0 &    0 &  0 &  0 \\

na+n3 & 0 &  0 &  0 &    0 &  0 &  0 \\

na+wr & 0 &  0 &  0 &    0 &  0 &  0 \\

na+n3m& 0 &  0 &  0 &    0 &  0 &  0 \\

na+n3e& 0 &  0 &  0 &    0 &  0 &  0 \\

na+n3g& 0 &  0 &  0 &    0 &  0 &  0 \\

na+giant & 0 &  0 &  0 &    0 &  0 &  0 \\
\hline

\end{tabular}
\end{table}

\begin{table}[h]
\caption[]{NGC5253B; age 5.0 Myrs; total mass $10^{6.6} M_{\odot}$}
\begin{tabular}{|l||c|c|c|c|c|c|}
\hline
 Slope&2.35&2.35&2.35& 1.35&1.35&1.35\\
\hline     

Up.mas.&120&  60&  40&   120&  60&  40\\

%\hline
\hline
bh+ms & 1 &  0 &  0 &   21&   0 &  0 \\

bh+giant & 0 &  0 &  0 &    1&   0 &  0 \\

bh+n3e& 1 &  0 &  0 &    7&   3 &  0 \\

bh+n3g& 2 &  1 &  0 &   36&   7 &  0 \\

bh+wr & 0 &  0 &  0 &    3&   0 &  0 \\

na+ms &11 & 10 &  5 &   92& 112 & 58 \\

na+n3 & 0 &  0 &  0 &    0&   0 &  0 \\

na+wr & 0 &  0 &  0 &    0&   0 &  0 \\

na+n3m& 0 &  0 &  0 &    0&   0 &  0 \\

na+n3e& 0 &  0 &  0 &    0&   0 &  0 \\

na+n3g& 0 &  0 &  0 &    0&   0 &  0 \\

na+giant & 0 &  0 &  0 &    0&   0 &  0 \\
\hline

\end{tabular}
\end{table}

\begin{table}[h]
\caption[]{Tol 89; age 4.5 Myrs; total mass $10^{5.7} M_{\odot}$}
\begin{tabular}{|l||c|c|c|c|c|c|}
\hline
 Slope&2.35&2.35&2.35& 1.35&1.35&1.35\\
\hline     

Up.mas.&120&  60&  40&   120&  60&  40\\

\hline
%\hline
bh+ms & 0 &  0 &  0 &    4 &  0 &  0 \\

bh+giant & 0 &  0 &  0 &    0 &  0 &  0 \\

bh+n3e& 0 &  0 &  0 &    1 &  0 &  0 \\

bh+n3g& 0 &  0 &  0 &    4 &  1 &  0 \\

bh+wr & 0 &  0 &  0 &    0 &  0 &  0 \\

na+ms & 1 &  1 &  0 &    9 &  9 &  2 \\

na+n3 & 0 &  0 &  0 &    0 &  0 &  0 \\

na+wr & 0 &  0 &  0 &    0 &  0 &  0 \\

na+n3m& 0 &  0 &  0 &    0 &  0 &  0 \\

na+n3e& 0 &  0 &  0 &    0 &  0 &  0 \\

na+n3g& 0 &  0 &  0 &    0 &  0 &  0 \\

na+giant & 0 &  0 &  0 &    0 &  0 &  0 \\

\hline
\end{tabular}
\end{table}

\begin{table}[h]
\caption[]{NGC3125; age 5.0 Myrs; total mass $10^{6.1} M_{\odot}$}
\begin{tabular}{|l||c|c|c|c|c|c|}
\hline
 Slope&2.35&2.35&2.35& 1.35&1.35&1.35\\
\hline

Up.mas.&120&  60&  40&   120&  60&  40\\

\hline     
%\hline
bh+ms & 0 &  0 &  0 &    7 &  0 &  0 \\

bh+giant & 0 &  0 &  0 &    0 &  0 &  0 \\

bh+n3e& 0 &  0 &  0 &    2 &  1 &  0 \\

bh+n3g& 1 &  0 &  0 &   12 &  2 &  0 \\

bh+wr & 0 &  0 &  0 &    1 &  0 &  0 \\

na+ms & 3 &  3 &  1 &   29 & 36 & 18 \\

na+n3 & 0 &  0 &  0 &    0 &  0 &  0 \\

na+wr & 0 &  0 &  0 &    0 &  0 &  0 \\

na+n3m& 0 &  0 &  0 &    0 &  0 &  0 \\

na+n3e& 0 &  0 &  0 &    0 &  0 &  0 \\

na+n3g& 0 &  0 &  0 &    0 &  0 &  0 \\

na+giant & 0 &  0 &  0 &    0 &  0 &  0 \\

\hline
\end{tabular}
\end{table}

Curves were not smoothed so all fluctuations of statistical nature are
presented. We calculated $10^7$ binary systems 
and then the results were normalized.

 We apply our results to seven regions of recent starformation (see the
tables, the full set can be found in (Popov et al., 1999)).
Ages, total masses and some other characteristics were taken from SCK98
(we used total masses determined for Salpeter's IMF even for the IMFs with
different parameters, which is a simplification).
We made an assumption, that binaries contain
50\% of the total mass of the starburst. Numbers were rounded off to the
nearest integer.

As far as for several regions ages are uncertain, we made calculations for 
two values of the age.

Different types of close binaries show different sensitivity to variations
of the IMF. When we replace $\alpha=2.35$ by $\alpha=1$ the numbers
of all sources increase. 
Systems with BHs are more sensitive to such variations. 

When one try to vary the upper mass limit, another situation appear.
In some cases (especially for $\alpha=2.35$) systems with NSs show
little differences for different values of the upper mass limit, 
while systems with BHs become significantly  less (or more) abundant
for different upper masses.
Luckily, X-ray transients, which are the most numerous systems in our
calculations,
show significant sensitivity to variations of the upper mass limit.   
But of course due to their transient nature it is difficult to
use them to detect small variations in the IMF.
If it is possible to distinguish systems with BH, it is much better to use
them to test the IMF.\\[3mm] 

{\bf 4. Discussion and conclusions}\\[1mm]

The results of our calculations can be easily used to estimate 
the number of X- ray sources for different parameters of the IMF
if the total mass of stars and age of a starburst are known 
(in (Popov et al., 1997, 1998)
analytical approximations for source numbers were given). 
And we estimate numbers of different
sources for several regions of recent starformation.

Here we tried to show, that
populations of close binaries are very sensitive
to the variations of the IMF. One must be careful,
when trying to fit the observed data for single stars
with  variations of the IMF. And, vice versa, using detailed observations of
X-ray sources, one can try to estimate parameters of the IMF, and test
results, obtained from single stars population.

{\it Acknowledgements.}  We want to thank K.A. Postnov for discussions and
G.V. Lipunova and I.E. Panchenko for technical assistance.
SBP also thanks organizers of the conference for 
support and hospitality.

 This work was supported by the grants: NTP ``Astronomy'' 1.4.2.3., 
NTP `Astronomy'' 1.4.4.1 and ``Universities of Russia'' N5559.\\[3mm]
\indent
{\bf References\\[2mm]}
Contini, T., Davoust, E., \& Considere, S.: 1995,  A \& A {\bf 303}, 440\\
Lipunov, V.M., Ozernoy, L.M., Popov, S.B., Postnov, K.A. \&
Prokhorov, M.E.: 1996a, {\it ApJ} {\bf 466}, 234\\
Lipunov, V.M., Postnov, K.A. \& Prokhorov, M.E.: 1996b, {\it Astroph.
and Space Phys. Rev.} {\bf 9}, part 4\\
Nazin, , S.N., Lipunov, V.M., Panchenko, I.E., Postnov, K.A.,
Prokhorov, M.E. \& Popov, S.B.: 1998, {\it Grav. \& Cosmology} {\bf 4}, 
suppl. ``Cosmoparticle Physics'' part.1, 150 (astro-ph 9605184)\\ 
Popov, S.B.,  Lipunov, V.M., Prokhorov, M.E., \&  Postnov, K.A.:
1997, {\it astro-ph/9711352}\\
Popov, S.B.,  Lipunov, V.M., Prokhorov, M.E., \&  Postnov, K.A.:
1998, {\it AZh}, {\bf 75}, 35 (astro-ph/9812416)\\ 
 Popov, S.B., Prokhorov, M.E.,  \& Lipunov, V.M.: 1999, {\it 
astro-ph/9905070}\\
Schaerer, D.: 1996, {\it ApJ} {\bf 467}, L17\\
Schaerer, D., Contini, T., \& Kunth, D.: 1998, {\it A\&A} {\bf 341}, 399 
(astro-ph/9809015) (SCK98)\\
van den Heuvel, E.P.J.: 1994, in ``Interacting Binaries'', 
Eds. Shore, S.N., Livio, M., \& van den Heuvel, E.P.J., Berlin, Springer,
442\\

\vfill
\end{document}